\documentclass{article}
\usepackage{spconf,amsmath,graphicx}
\usepackage[detect-all]{siunitx}
\usepackage{mathtools}
\usepackage{comment}
\usepackage{multirow}
\usepackage{bbm}
\usepackage[subrefformat=parens]{subfig}
\usepackage{threeparttable}
\usepackage{booktabs}
\usepackage{amssymb}
\usepackage{hyperref}

\usepackage{xspace}
\makeatletter
\DeclareRobustCommand\onedot{\futurelet\@let@token\@onedot}
\def\@onedot{\ifx\@let@token.\else.\null\fi\xspace}
\def\eg{\emph{e.g}\onedot} 
\def\ie{\emph{i.e}\onedot}

\makeatother

\def\equationautorefname~#1\null{(#1\null)}

\newcommand{\vect}[1]{\mbox{\boldmath $#1$}}

\newcommand{\norm}[1]{\left\lVert#1\right\rVert}

\newcommand{\trans}[1]{#1^\mathsf{T}}

\DeclareMathOperator*{\argmin}{arg\,min}\def\appendixautorefname~#1\null{~#1 \null}

\newcommand{\tablescale}{0.8}

\title{Towards Neural Diarization for Unlimited Numbers of Speakers\\Using Global and Local Attractors}
\name{Shota Horiguchi$^1$, Shinji Watanabe$^2$, Paola Garc\'{i}a$^3$, Yawen Xue$^1$, Yuki Takashima$^1$, Yohei Kawaguchi$^1$}
\address{
  $^1$Hitachi, Ltd., Japan\\
  $^2$Carnegie Mellon University, USA\\
  $^3$Johns Hopkins University, USA}

\begin{document}
\ninept
\abovedisplayskip=4pt
\belowdisplayskip=4pt

\setlength\floatsep{10pt}
\setlength\textfloatsep{10pt}
\setlength\intextsep{10pt}
\setlength\abovecaptionskip{4pt}
\setlength\belowcaptionskip{4pt}

\aboverulesep=0.25ex 
\belowrulesep=0.5ex 

\maketitle
\begin{abstract}
    Attractor-based end-to-end diarization is achieving comparable accuracy to the carefully tuned conventional clustering-based methods on challenging datasets.
    However, the main drawback is that it cannot deal with the case where the number of speakers is larger than the one observed during training.
    This is because its speaker counting relies on supervised learning.
    In this work, we introduce an unsupervised clustering process embedded in the attractor-based end-to-end diarization.
    We first split a sequence of frame-wise embeddings into short subsequences and then perform attractor-based diarization for each subsequence. 
    Given subsequence-wise diarization results, inter-subsequence speaker correspondence is obtained by unsupervised clustering of the vectors computed from the attractors from all the subsequences.
    This makes it possible to produce diarization results of a large number of speakers for the whole recording even if the number of output speakers for each subsequence is limited.
    Experimental results showed that our method could produce accurate diarization results of an unseen number of speakers.
    Our method achieved \SI{11.84}{\percent}, \SI{28.33}{\percent}, and \SI{19.49}{\percent} on the CALLHOME, DIHARD II, and DIHARD III datasets, respectively, each of which is better than the conventional end-to-end diarization methods.
\end{abstract}

\section{Introduction}
Analyzing who is speaking when in a multi-talker conversation plays an essential role in many speech-related applications.
This task is called speaker diarization, and in recent years there have been extensive efforts toward accurate speaker diarization in various domains \cite{park2021review}.
Although the conventional methods based on clustering of speaker embeddings are still powerful baselines \cite{landini2022bayesian}, end-to-end methods are almost reaching their performance (see the results of the DIHARD III challenge \cite{ryant2021third}), thanks largely to their way of handling overlapped speech.

While some end-to-end methods are reaching the accuracy of conventional clustering-based approaches that utilize speaker embeddings, they still have difficulty in correctly estimating the number of speakers.
For clustering-based methods, the number of speakers is determined in the clustering step, making the methods flexible.
On the other hand, some end-to-end methods fix the number of speakers \cite{fujita2019end2,medennikov2020targetspeaker,liu2021endtoend}.
This assumption does not matter in some fixed-domain scenarios (\eg, two-speaker telephone conversations \cite{callhome}, doctor-patient conversations \cite{shafey2019joint}, or four-speaker dinner parties \cite{watanabe2020chime}), but becomes a crucial problem in the harder scenarios \cite{ryant2021third}.
There are a few recent methods that can handle a flexible number of speakers \cite{horiguchi2020endtoend,takashima2021endtoend,maiti2021endtoend}, but since they are trained in a fully supervised manner, they still face the problem of not being able to produce a larger number of speakers than that observed during training.
A solution to the mismatch in the number of speakers has not been established yet.

To tackle this problem, this paper introduces a clustering step into the attractor-based end-to-end neural diarization (EEND-EDA) \cite{horiguchi2020endtoend,han2021bwedaeend,horiguchi2021encoderdecoder}.
We first clarify that the maximum number of speakers that can be estimated stems from the attractor calculation part; the embedding extraction part based on stacked Transformer encoders is well generalized to an unseen number of speakers.
To extend EEND-EDA to an unlimited number of speakers, we focus on the fact that the number of speakers who speak in a short period is sufficiently small \cite{yoshioka2019advances,chen2020continuous}.
Therefore, given a sequence of frame-wise embeddings, we first split them into short subsequences and calculate speaker-wise local attractors and speech activities based on them for each subsequence.
Then, the system finds inter-subsequence attractor correspondence by clustering the vectors converted from the local attractors.
The number of clusters, \ie, the number of speakers, is automatically decided by eigenvalue analysis of the affinity matrix calculated based on the local attractors.
Because the number of speakers observed in each subsequence can be regarded as small, we can still use the attractor calculation part.
We show that the proposed method can produce diarization results for a large number of speakers different from the number of speakers in the training data.
We also show that it achieves comparable performance to conventional methods including carefully tuned challenge submissions.

\section{EEND-EDA and its limitation}
\begin{figure*}[t]
    \centering
    \begin{minipage}{0.37\linewidth}
        \centering
        \includegraphics[width=0.49\linewidth]{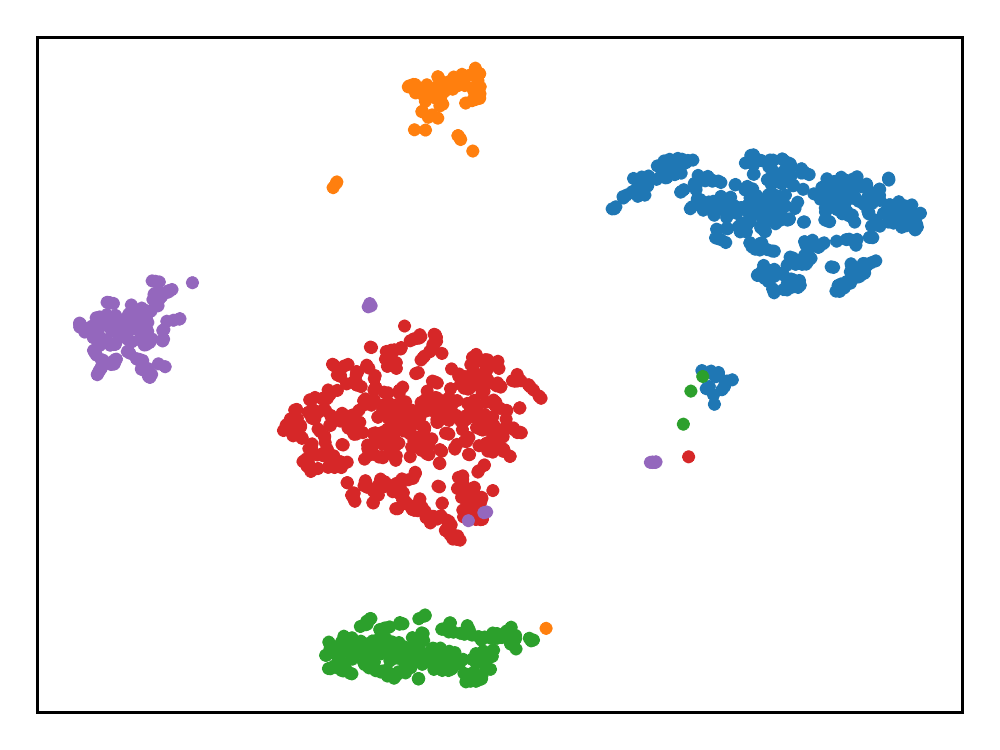}
        \includegraphics[width=0.49\linewidth]{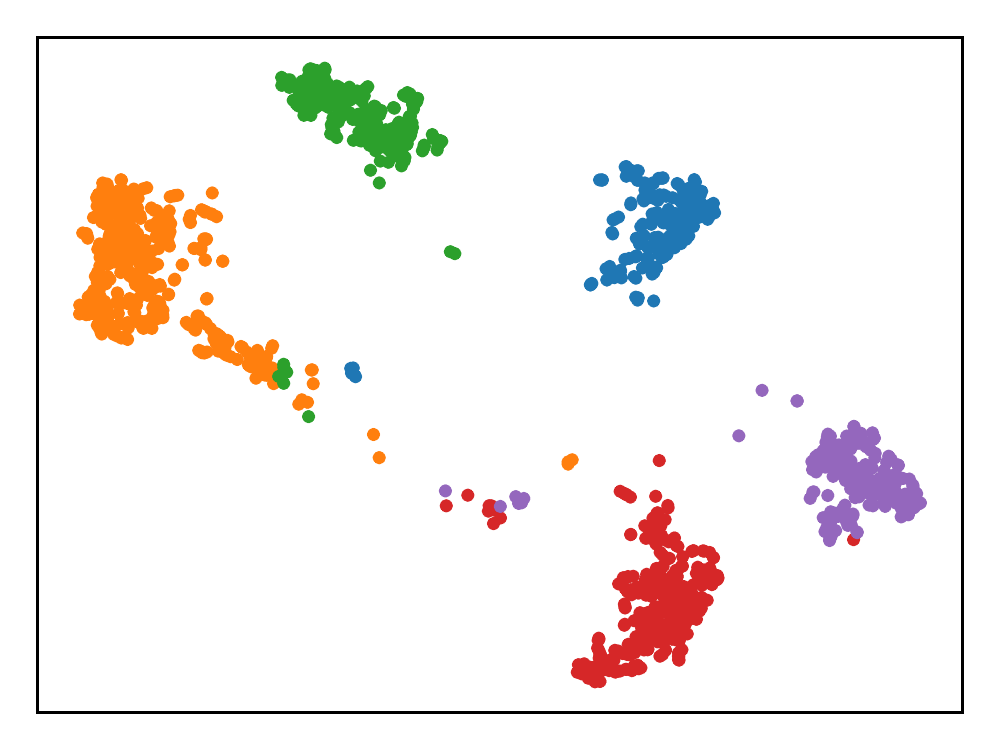}\\
        \includegraphics[width=0.49\linewidth]{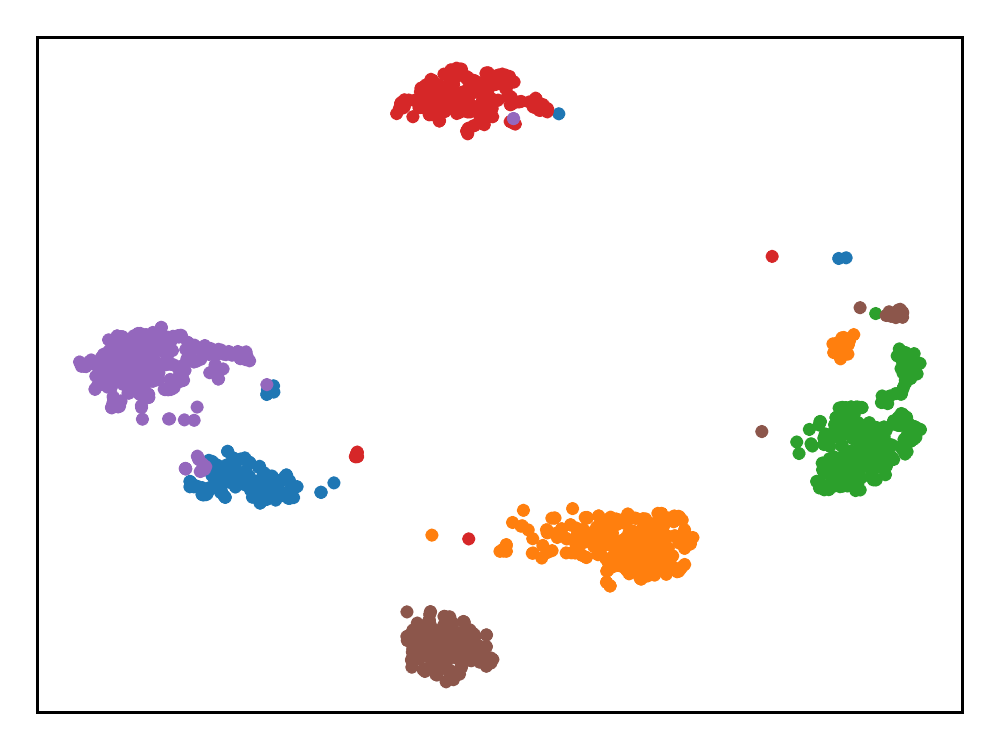}
        \includegraphics[width=0.49\linewidth]{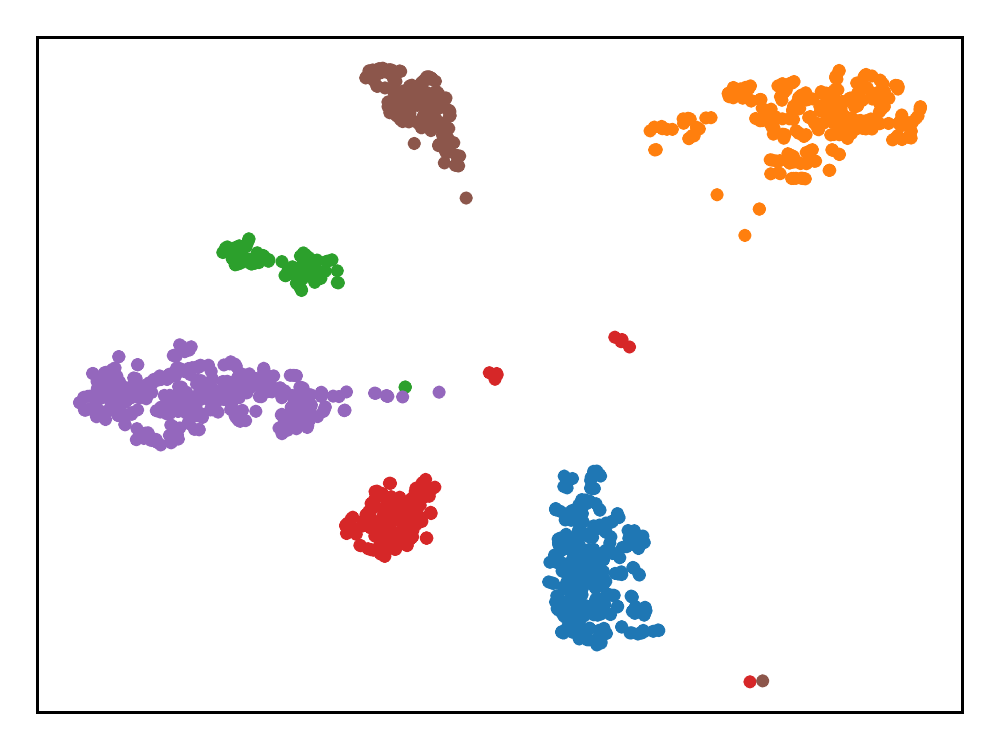}\\
        \caption{Visualization of frame-wise embeddings of simulated 5-speaker mixtures (top) and 6-speaker mixtures (bottom). The embeddings were extracted using the model trained on \{1,2,3,4\}-speaker mixtures. The color of each dot corresponds to the speaker identity. Overlapped frames are excluded from the visualization.}
        \label{fig:visualization}
    \end{minipage}
    \hfill
    \begin{minipage}{0.6\linewidth}
        \centering
        \includegraphics[width=\linewidth]{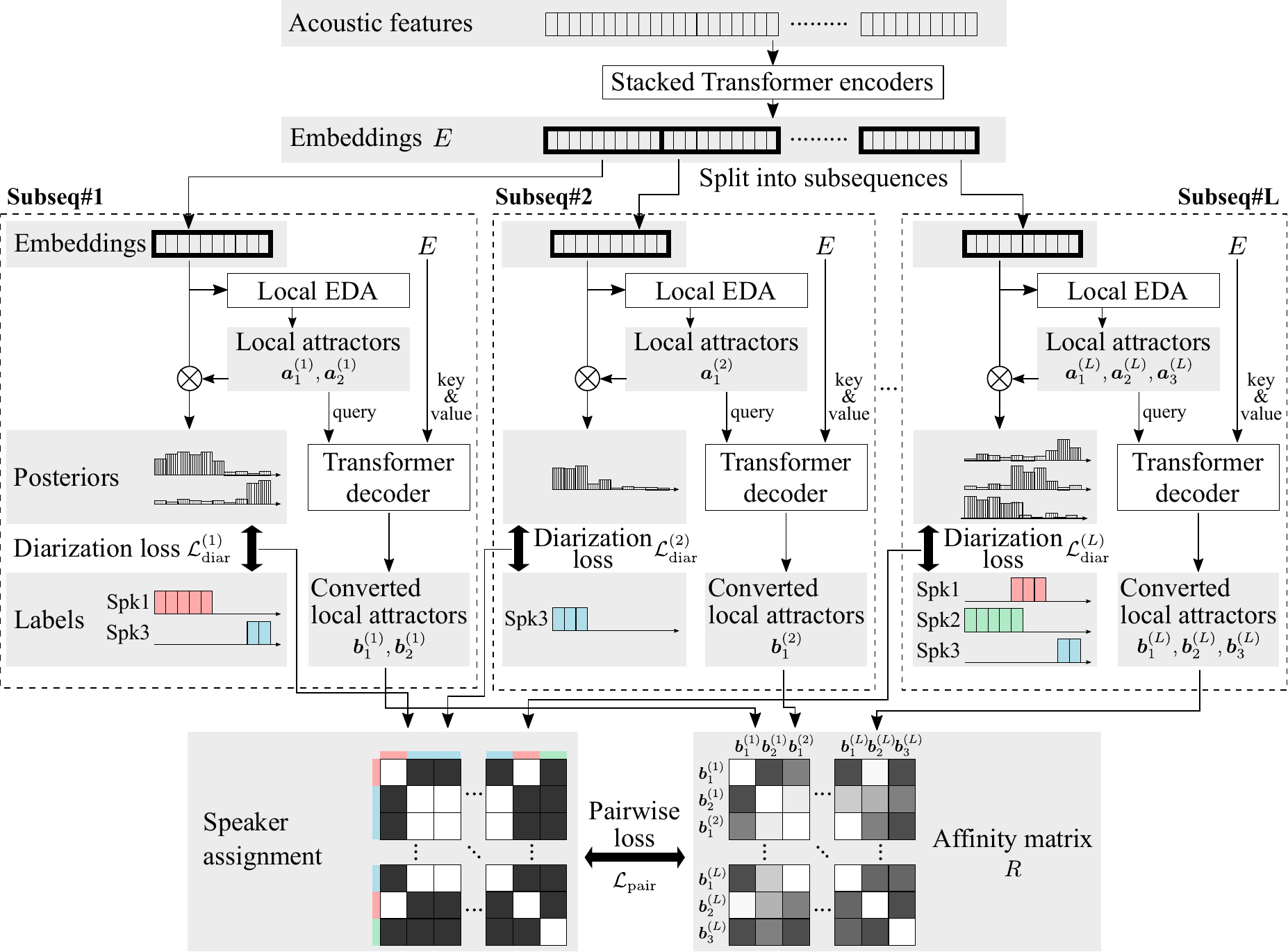}
        \caption{EEND based on local attractors.}
        \label{fig:diagram}
    \end{minipage}
\end{figure*}
\subsection{Brief introduction to EEND-EDA}\label{sec:eend_eda}
EEND-EDA \cite{horiguchi2020endtoend,horiguchi2021encoderdecoder} is an end-to-end trainable diarization model that can treat a flexible number of speakers' overlapping speech.
Given a $T$-length sequence of $F$-dimensional acoustic features $\left(\vect{x}_t\right)_{t=1}^T$, EEND-EDA first converts them into the same length of $D$-dimensional embeddings $\left(\vect{e}_t\right)_{t=1}^T$ using stacked Transformer encoders without positional encoding.
To calculate speaker-wise flexible number of attractors $\left\{\vect{a}_s\in(-1,1)^D\right\}_s$, where $s$ is the speaker index, the embeddings are passed through the encoder-decoder-based attractor calculation module (EDA):
\begin{align}
    \vect{a}_1,\vect{a}_2,\dots=\mathsf{EDA}\left(\vect{e}_1,\dots,\vect{e}_T\right).
    \label{eq:eda}
\end{align}
In this paper, we refer to these attractors calculated from a whole recording as \textit{global attractors}.
We stop the attractor calculation if the attractor existence probability
\begin{align}
    \hat{z}_s=\sigma\left(\trans{\vect{w}}\vect{a}_s+b\right)\in\left(0,1\right)\quad\left(s=1,2,\dots\right)
    \label{eq:attractor_existence_probability}
\end{align}
is below 0.5, where $\sigma\left(\cdot\right)$ is the sigmoid function and $\vect{w}\in\mathbb{R}^D$ and $b\in\mathbb{R}$ are the trainable weights and bias of a fully connected layer, respectively.
During training, we know the oracle number of speakers $S$ so that $S+1$ attractors are calculated to compute the attractor existence loss $\mathcal{L}_\text{exist}$:
\begin{align}
    \mathcal{L}_\text{exist}=\frac{1}{S+1}\sum_{s=1}^{S+1}H\left(z_{s},\hat{z}_{s}\right),\quad
    z_s=\begin{cases}
        1&(s\in\{1,\dots,S\})\\
        0&(s=S+1)
    \end{cases},
    \label{eq:loss_att}
\end{align}
where $H\left(z_s,\hat{z}_s\right)$ is the binary cross entropy determined as
\begin{align}
    H\left(z_s,\hat{z}_s\right)=-z_s\log{\hat{z}_s}-\left(1-z_s\right)\log\left(1-\hat{z}_s\right).
\end{align}
The estimation of speaker $s$'s speech activity at $t$ is calculated as the dot product of the corresponding embedding and attractor, as
\begin{align}
    \hat{y}_{t,s}=\sigma\left(\trans{\vect{e}_t}\vect{a}_s\right)\in\left(0,1\right).
    \label{eq:diarization}
\end{align}
The posterior $\hat{y}_{t,s}$ is optimized using the following diarization loss:
\begin{align}
    \mathcal{L}_\text{diar}=\frac{1}{TS}\argmin_{\left(\phi_1,\dots,\phi_T\right)\in\Phi\left(1,\dots,T\right)}\sum_{t=1}^T\sum_{s=1}^{S}H\left(y_{\phi_t,s},\hat{y}_{t,s}\right),
    \label{eq:loss_diar}
\end{align}
where $\Phi\left(1,\dots,T\right)$ is a set of all the permutations of $\left(1,\dots,T\right)$ and $y_{t,s}\in\{0,1\}$ is the ground-truth label of speaker $s$'s speech activity at $t$.
The total loss $\mathcal{L}_\text{global}$ is the weighted sum of the diarization loss and attractor existence loss:
\begin{align}
    \mathcal{L}_\text{global}=\mathcal{L}_\text{diar}+\alpha\mathcal{L}_\text{exist},
    \label{eq:loss_global}
\end{align}
where $\alpha$ is the weighting parameter; we set $\alpha=1$ in this paper.
Note that the loss $\mathcal{L}_\text{exist}$ is used to update only $\vect{w}$ and $b$ in \autoref{eq:attractor_existence_probability}, which was found to contribute to improve the performance \cite{horiguchi2021encoderdecoder}, while the original paper \cite{horiguchi2020endtoend} used it to update the parameters of the Transformer encoders and LSTM encoder-decoder as well.

\subsection{Limitation of EEND-EDA}
EEND-EDA can handle recordings of a flexible number of speakers, but it is empirically known that the number of output speakers is capped by that in the training dataset (see \autoref{sec:result_simu}).
Even if it is adapted to datasets that contain more speakers than four (\eg, DIHARD \cite{ryant2019second,ryant2021third}), it is hard to produce diarization results of more than four speakers.

\autoref{fig:visualization} shows the t-SNE visualization of the frame-wise embeddings $\vect{e}_t$ using five- and six-speaker mixtures.
Here, EEND-EDA was trained on mixtures that each contain four speakers at most.
We can clearly see that the speakers are well-separated in the embedding space even when the input mixture contains a larger number of speakers than the training data.
This indicates that the number of output speakers is limited owing to the EDA operation in \autoref{eq:eda}, despite the speaker separation capabilities of the EEND embedding vectors $\vect{e}_t$.
In the next section, we explain how this problem is solved in the proposed method.

\section{Proposed method: Neural diarization based on global and local attractors}
\subsection{Overview}
One possible way to ease this limitation is to train the model using mixtures that contain large numbers of speakers \cite{maiti2021endtoend}.
However, it is difficult to increase the number of speakers inexhaustibly.
This is because the EEND's training procedure depends on a permutation-free objective \cite{hershey2016deep,yu2017permutation}, which takes $\mathcal{O}\left(TS^2\right)+\mathcal{O}\left(S^3\right)$ operation even with the optimal mapping loss \cite{lin2020optimal}.
It is also a problem that EEND uses a fixed length of chunks (\eg 500 frames) for efficient batch processing during training, and it is rare to contain a large number of speakers within such a chunk even if the whole recording can.
Moreover, we do not know how many speakers would be sufficient for training.
Therefore, we have to treat the problem without using mixtures with a large number of speakers.

In this paper, we assume that the number of speakers in a short period is small \cite{yoshioka2019advances,chen2020continuous}.
We apply EDA for each short subsequence to calculate attractors called \textit{local attractors}, and then cluster the local attractors to find inter-subsequence speaker correspondence.
Even though the number of speakers for each subsequence is limited, the total number of speakers can be larger than the limitation.

\subsection{Training}
The schematic diagram of the proposed local-attractor-based diarization is shown in \autoref{fig:diagram}.
As introduced in \autoref{sec:eend_eda}, given the embeddings $\left(\vect{e}_t\right)_{t=1}^T\eqqcolon E$, which are output from the stacked Transformer encoders, we first split them into multiple short length $L$ subsequences $\left\{\left(\vect{e}_t\right)_{t=t_{l-1}+1}^{t_{l}}\right\}_{l=1}^{L}$, where $0=t_0<\dots<t_L=T$.
From the $l$-th sequence, we calculate local attractors $\left[\vect{a}_1^{(l)},\dots,\vect{a}_{S_l}^{(l)}\right]\eqqcolon A_l\in\left(-1,1\right)^{D\times S_l}$ by using \autoref{eq:eda}, where $S_l$ is the number of speakers active during $t_{l-1}+1\leq t\leq t_l$.
Note that the local attractors are calculated only for active speakers during each subsequence; thus, even if the $S$ speakers appeared during $T$ frames, the number of speakers in each sequence may be smaller than $S$, \ie, $0\leq S_l\leq S$.
The diarization loss $\mathcal{L}_\text{diar}^{(l)}$ and attractor existence loss $\mathcal{L}_\text{exist}^{(l)}$ are calculated for each subsequence by using \autoref{eq:loss_diar} and \autoref{eq:loss_att}, respectively.

With the two losses, we can estimate the number of speakers and speech activities for that number of speakers for each subsequence.
Here, the problem is how to find whether a pair of attractors from two subsequences correspond to the same speaker or different speakers.
To optimize the attractor distribution, we define the training objective based on the contrastive loss.

Because the local attractors are optimized to minimize the diarization error, we convert them to be suitable for clustering.
Given local attractors and frame-wise embeddings for each subsequence, we first convert the attractors using a Transformer decoder:
\begin{align}
    \vect{b}_1^{(l)},\dots,\vect{b}_{S_l}^{(l)}=\mathsf{TransformerDecoder}\left(A_l,E\right).
\end{align}
Here, the local attractors $A_l$ are queries and the embeddings $E$ are keys and values input to the Transformer decoder.
With the converted local attractors $B_l\coloneqq\left[\vect{b}_1^{(l)},\dots,\vect{b}_{S_l}^{(l)}\right]\in\mathbb{R}^{D\times S_l}$, the contrastive loss is calculated on the converted vectors from all the subsequences $B=\left[\vect{b}_i\right]_i\coloneqq\left[B_1,\dots,B_L\right]\in\mathbb{R}^{D\times S^*}$, where $S^*\coloneqq\sum_{l=1}^LS_l$.
When we calculate the subsequence-wise diarization loss $\mathcal{L}_\text{diar}^{(l)}$ using \autoref{eq:loss_diar}, we find the optimal mapping between the estimated and ground-truth speakers, \ie, we know whether the $i$-th and $j$-th local attractors (or converted local attractors) correspond to the same speaker.
Thus, the pairwise loss can be calculated for each pair of the converted local attractors as follows:
\begin{align}
    \mathcal{L}_\text{pair}=\sum_{i,j\in\{1,\dots,S^*\}}&\frac{1}{S^2c_ic_j}\left(r_{ij}\left(1-\operatorname{sim}\left(\vect{b}_i,\vect{b}_j\right)\right)\right. +\nonumber\\
    &\left.\left(1-r_{ij}\right)\left[\operatorname{sim}\left(\vect{b}_i,\vect{b}_j\right)-\delta\right]_{+}\right),\label{eq:pairwise_loss}
\end{align}
where $c_i$ ($c_j$) is the number of attractors that correspond to the $i$-th ($j$-th) attractor's speaker, $\operatorname{sim}\left(\vect{b}_i,\vect{b}_j\right)\coloneqq\frac{\trans{\vect{b}_i}\vect{b}_j}{\norm{\vect{b}_i}\norm{\vect{b}_j}}$ is the cosine similarity between $\vect{b}_i$ and $\vect{b}_j$, $r_{ij}$ is the indicator that takes $1$ if $\vect{b}_i$ and $\vect{b}_j$ correspond to the same speaker and $0$ otherwise, and $\left[\cdot\right]_{+}$ is the hinge function.
This pairwise loss aims to make the angle between converted attractors of the same speaker be zero and those of different speakers be at least $\arccos{\delta}$ apart.
In this paper, $\delta$ is set to 0.5 during pretraining and to 0 during adaptation.
Note that this loss is highly influenced by the instance segmentation in computer vision \cite{fathi2017semantic,kong2018recurrent}.
The operation of grouping pixel-wise embeddings calculated from a single image into instances is very similar to the current problem of grouping multiple local attractors from a single input into speaker identities.
X-vectors or frame-wise embeddings (\eg, $\left(\vect{e}_t\right)_{t=1}^T$) cannot be divided by speakers because natural conversations include overlaps.
However, each local attractor here corresponds to one speaker so that each local attractor can be hardly assigned to one of the clusters.

During the training, the following loss is used instead of \autoref{eq:loss_global} to optimize the diarization error within each subsequence and the distribution of the local attractors across subsequences:
\begin{align}
    \mathcal{L}_\text{local}=\frac{1}{L}\sum_{l=1}^L\left(\mathcal{L}_\text{diar}^{(l)}+\alpha\mathcal{L}_\text{exist}^{(l)}\right)+\gamma\mathcal{L}_\text{pair},
    \label{eq:local_loss}
\end{align}
where $\gamma$ is the weighting parameter, which is set to 1 in this paper.

We found that the model training that fully relies on the local-attractor-based loss \autoref{eq:local_loss} resulted in slow and unstable convergence.
To make use of global consistency, we also use the loss that utilizes both local and global attractors, defined as
\begin{align}
    \mathcal{L}_\text{both}=\mathcal{L}_\text{local}+\mathcal{L}_\text{global}.
    \label{eq:both_loss}
\end{align}

\subsection{Inference}
In the inference phase, we first estimate the number of speakers $\hat{S}_l\in\mathbb{Z}_{\geq0}$ using attractor existence probabilities in \autoref{eq:attractor_existence_probability}, and then estimate speech activities $\hat{\vect{y}}_l\in\left(0,1\right)^{\hat{S}_l\times\left(t_l-t_{l-1}\right)}$ from each subsequence by using \autoref{eq:diarization}.
We then apply unsupervised clustering for the converted local attractors from all the subsequences.
The crucial problem here is how to define the clustering parameters, \eg, the number of clusters or threshold values.

One approach consists of the following processes: 1) construct an affinity matrix, 2) calculate its graph Laplacian, 3) conduct eigenvalue decomposition, and 4) estimate the number of speakers based on the maximum eigengap.
Because the affinity matrix calculated from speaker embeddings often contains unreliable values, it is important to remove noises from it.
For example, Gaussian blur was applied to smooth the affinity matrix calculated from d-vectors extracted using sliding window \cite{wang2018speaker}.
For our local-attractor-based method, however, smoothing cannot be used because attractors are calculated not only for each subsequence but also for each speaker within a subsequence.
In \cite{park2020auto}, $p$ nearest neighbor binarization was applied to the affinity matrix to remove unreliable values.
The value of $p$ is selected automatically, and empirically dozens of nearest neighbors are used.
In our case, however, this is not suitable because local attractors are calculated every five seconds in this paper, so the number of vectors to be clustered is extremely insufficient.

Let us consider why the noise inhibits the accurate approximation of the number of clusters in the eigengap-based estimation.
One reason is that the eigenvalues of a graph Laplacian are obtained without considering the size of clusters; thus, noises produce a lot of tiny clusters.
To penalize more on small clusters, we directly use an affinity matrix instead of its graph Laplacian.
Given the affinity matrix $R=\left(r_{ij}\right)\in\left[-1,1\right]^{S^*\times S^*}$, where $r_{ij}=\mathrm{sim}\left(\vect{b}_i,\vect{b}_j\right)$, we apply matrix decomposition as
\begin{align}
    \begingroup
    \setlength\arraycolsep{2pt}
    R=V\begin{bmatrix}
        \lambda_1&\\[-6pt]
        &\ddots&\\[-4pt]
        &&\lambda_{S^*}\\
    \end{bmatrix} V^{-1},
    \endgroup
    \label{eq:matrix_decomposition}
\end{align}
where $V\in\mathbb{R}^{S^*\times S^*}$ is the eigenvectors and $\lambda_1>\dots>\lambda_{S^*}$ are the eigenvalues of $R$.
Because $R$ is positive-semidefinite, the eigenvalues are non-negative, and they indicate the size of each cluster where each local attractor is softly assigned.
The number of clusters can be estimated by using the eigenratio instead of eigengap, as
\begin{align}
    \hat{S}=\min_{1\leq s\leq S^*-1}\frac{\lambda_{s+1}}{\lambda_{s}}.
\end{align}
In this paper, we use the hinge function in \autoref{eq:pairwise_loss}.
We also know that the attractors from the same subsequence must be assigned to different clusters.
Therefore, we use the modified affinity matrix $R'=\left(r'_{ij}\right)\in\left[0,1\right]^{S^*\times S^*}$ instead of $R$, defined as
\begin{align}
    r'_{ij}=\begin{cases}
    \mathbbm{1}\left(i=j\right)&\left(\text{$\vect{b}_i$ and $\vect{b}_j$ are from}\right.\\
    &\left.\text{the same subsequence}\right)\\
    \frac{1}{1-\delta}\left[\operatorname{sim}\left(\vect{b}_i,\vect{b}_j\right)-\delta\right]_{+}&(\text{otherwise})
    \end{cases},
\end{align}
where $\mathbbm{1}\left(\mathrm{cond}\right)$ is an indicator function that returns $1$ if $\mathrm{cond}$ is true and $0$ otherwise.
We then apply matrix decomposition as in \autoref{eq:matrix_decomposition} and obtain eigenvalues $\lambda'_1\geq\dots\geq\lambda'_{S^*}$.
Indeed, $R'$ is no longer positive-semidefinite but its eigenvalues are still good indicators of the size of clusters.
Because the eigenvalues indicate the size of clusters, we only use those of not less than one to estimate the number of speakers $\hat{S'}$ as
\begin{align}
    \hat{S'}=\min_{\substack{1\leq s\leq S^*-1 \\ \lambda'_s\geq1}}\frac{\lambda'_{s+1}}{\lambda'_{s}}.
\end{align}

Even if we force the affinity value between a pair of local attractors from the same subsequence to be zero in $R'$, the local attractors may belong to the same cluster; thus, the number of estimated speakers can be less than the maximum number of speakers from one of the subsequences.
Therefore, we update the estimation of the number of speakers by
\begin{align}
    \hat{S'}\leftarrow\max\left(\hat{S'},\max_{1\leq l\leq L} S_l\right).
\end{align}

After the number of speakers is estimated, we apply a clustering method to the local attractors.
We know that the attractors from the same subsequence have to be assigned to different clusters, so it is effective to use cannot-link constraints for clustering.
One possible choice is to use COP-Kmeans clustering \cite{wagstaff2001constrained}, which is used in EEND-vector clustering \cite{kinoshita2021integrating,kinoshita2021advances}, but this has a difficulty in the case where cannot-link constraints have to be satisfied; the algorithm sometimes results in no solution.
Thus, we used the CLC-Kmeans algorithm \cite{yang2013improved} instead for stable convergence.

The model that is trained using \autoref{eq:both_loss} also has outputs based on the global branch.
They are still useful when the number of clusters is small because the model is trained in a fully supervised manner.
Therefore, we propose using global- and local-attractor-based inference depending on the estimated number of speakers.
We first estimate the number of speakers by using \autoref{eq:attractor_existence_probability} calculated from global attractors.
In this paper, we trained the model using \{1,2,3,4\}-speaker mixtures; thus, if the number of estimated speakers is less than four, we use inference based on global attractors explained in \autoref{sec:eend_eda}. If the estimated number of speakers is equal to or larger than four, we use inference based on local attractors explained in this section.
We call this inference switching strategy.

\section{Related works}
\subsection{Speaker diarization}
The clustering-based methods generally consist of the following: speech activity detection (SAD), speaker embedding extraction, clustering of the embeddings, and optional overlap assignment.
The SAD is often replaced by the oracle speech segments, but the remaining parts are actively being studied, \eg, investigation of better architectures for speaker embedding extractors \cite{zhou2021resnext}, better clustering methods \cite{landini2022bayesian,zhang2019fully,li2021discriminative}, and better overlap assignment methods \cite{bullock2020overlap}.
An important property of the clustering-based methods is that they do not limit the number of speakers that can be estimated during inference because the results are obtained by unsupervised clustering.

On the other hand, neural-network-based diarization methods are rapidly emerging to replace the conventional clustering-based methods, but they still have a limitation.
For example, personal VAD \cite{ding2020personal} and VoiceFilter-Lite \cite{wang2020voicefilterlite} assume that the target speaker's d-vector is available during inference, so they are not suitable for speaker-independent diarization.
Target-speaker voice activity detection \cite{medennikov2020targetspeaker} and the initial models of end-to-end neural diarization (EEND) \cite{fujita2019end2} fix the output number of speakers, so they are not suitable for diarization of unknown numbers of speakers.
The recurrent selective attention network (RSAN) \cite{kinoshita2020tackling} or some extensions of EEND \cite{horiguchi2020endtoend,takashima2021endtoend} can deal with flexible numbers of speakers.
However, EEND-based models empirically limit the number of output speakers by the number of speakers in the training datasets. It is unclear whether the RSAN can deal with the number of speakers limitation because only the speaker counting accuracy on matched conditions (zero, one, or two speakers) has been reported.
EEND as post-processing \cite{horiguchi2021endtoend} tackled the problem of the limitation of the number of speakers by taking advantage of both clustering-based methods and end-to-end methods through the utilization of EEND to refine the results from clustering, but a single model solution for this problem is still awaited.

The most relevant work is the recently proposed EEND-vector clustering, which incorporates EEND and speaker embeddings \cite{kinoshita2021integrating,kinoshita2021advances}, but it differs from ours in a few key ways.
One is that EEND-vector clustering relies on the speaker embedding dictionary, which requires speaker identities across recordings in the training set.
Such information is accessible in the simulated mixtures created from single-speaker recordings (\eg, NIST SRE, Librispeech \cite{panayotov2015librispeech}, VoxCeleb \cite{nagrani2020voxceleb}), but is not always available in real conversation datasets (\eg, DIHARD \cite{ryant2019second,ryant2021third}).
Our method only utilizes the speaker information within each recording so that it can use such datasets for training.
Another difference is that EEND-vector clustering calculates speaker embeddings for each block, and it has been reported in \cite{kinoshita2021advances} that the block size needs to be somewhat long (\eg. \SI{30}{\second}) to obtain reliable speaker embeddings.
However, since the number of speakers that can appear in a block is limited by the network architecture, increasing the block size also limits the number of speakers that can appear in the final diarization results.
In contrast, in our method, we process a sequence of acoustic features by means of a stacked Transformer encoder before splitting them into subsequences.
Therefore, the frame-wise embeddings $\vect{e}_t$ can capture global context, and thus, we can use a shorter size of subsequence (\SI{5}{\second} in this paper) than EEND-vector clustering.

\subsection{Efforts to produce results for larger number of speakers than that observed during training}
Supervised speech processing methods sometimes suffer from the number of speakers mismatch between training and inference, especially when more speakers appear during inference than during training.
Some neural-network-based speech separation methods \cite{yu2017permutation,wang2018alternative,luo2018tasnet,luo2019conv,nachmani2020voice} limit the number of outputs by their network architecture, and thus, there is no way to deal with the mismatch.
Even if the method itself is designed not to limit the number of speakers, there is rarely experimental evidence to show how the model actually works under mismatched conditions \cite{chen2017deep,zeghidour2021wavesplit}.
In terms of speaker diarization, some EEND-based models \cite{horiguchi2020endtoend,takashima2021endtoend} can deal with flexible numbers of speakers, but the mismatch of the number of speakers remains an open question.
In this subsection, we introduce two successful speech processing approaches for the mismatch.

The first approach is one-and-rest permutation invariant training (OR-PIT) \cite{takahashi2019recursive}, which aims to split a mixture into the one-speaker waveform and the mixture of the rest of the speakers as residual output.
Even though the model is trained only on two- and three-speaker mixtures, the experimental results demonstrated that it worked well on four-speaker mixtures.
To adopt the one-vs-rest approach, it is necessary to decide the residual output.
In the context of speech separation, this can be easily determined by waveforms or time-frequency masks.
However, in the context of diarization, we cannot determine such residual output because we cannot assume the maximum number of sources for each frame; thus, it cannot be used for diarization.

The second approach is to introduce unsupervised clustering into the decoding step.
Deep clustering paper \cite{hershey2016deep} reported that time-frequency-bin-wise embeddings somewhat worked for three-speaker mixtures even if the model was only trained on two-speaker mixtures.
However, the number of speakers was assumed to be given and it is unclear whether we can share the same clustering parameters for matched and mismatched conditions.

\section{Experiments}
\subsection{Settings}
To train the proposed model, we created simulated multi-talker recordings using NIST SRE and Switchboard corpora following \cite{fujita2019end2}.
The average silence duration $\beta$ was varied to get a similar overlap ratio for each number of speakers, as shown in \autoref{tbl:simulated_dataset}.
For training, we created \{1,2,3,4\}-speaker simulated mixtures.
In addition, we created 5- and 6-speaker mixtures for evaluation to show that the proposed method can deal with a larger number of speakers than that observed during training.
See \cite{fujita2019end2} for the detailed protocol.

We also used real recordings summarized in \autoref{tbl:real_dataset} for evaluation.
For CALLHOME, we used Part 1 for model adaptation and Part 2 for testing. For DIHARD II and III, we used the development set for adaptation and the evaluation set for testing.

The original EEND-EDA \cite{horiguchi2020endtoend} was firstly trained on 2-speaker mixtures and then finetuned on \{1,2,3,4\}-speaker mixtures.
We found that training with subsequences from scratch resulted in poor model performance, so we first trained the model using 2-speaker mixtures with global attractors for 100 epochs and then finetuned it on \{1,2,3,4\}-speaker mixtures with the proposed method for another 50 epochs.
In the real dataset evaluations, we adapted the model for an additional 100 epochs on each adaptation set.
The Adam optimizer \cite{kingma2015adam} was used during training, with the Noam scheduler \cite{vaswani2017attention} with 100,000 warm-up steps for simulation-dataset-based training and a fixed learning rate of $1\times10^{-5}$ for adaptation.

As Transformer encoders, we used four-stacked encoders, which align with the experimental settings in the EEND-EDA papers \cite{horiguchi2020endtoend,horiguchi2021encoderdecoder}.
We also used six-stacked encoders with eight attention heads following the setting in the EEND-vector clustering paper \cite{kinoshita2021advances}.
As input features for the encoders, we used 345-dimensional log-mel filterbank-based acoustic features obtained every \SI{100}{\ms} following \cite{fujita2019end2,horiguchi2020endtoend}.
During training, the length of a sequence was set to \SI{50}{\second}, \ie, $L=500$, and the length of a subsequence was set to \SI{5}{\second}, \ie, $t_l=50l$ for $l\in\left\{0,\dots,10\right\}$.
During inference, each whole recording was processed at once and the length of each subsequence was set to \SI{5}{\second}.
To obtain high-resolution results for the DIHARD datasets, we used acoustic features extracted every \SI{50}{\ms}.

For evaluation metrics, we used diarization error rates (DERs) and Jaccard error rates (JERs).
Following prior studies \cite{fujita2019end2,horiguchi2020endtoend}, we allowed the collar tolerance of \SI{0.25}{\second} in the evaluations using the simulated datasets and the CALLHOME dataset, while we did not allow such collar in the evaluation of the DIHARD datasets.
Note that we did \textit{not} exclude overlapped speech from the evaluation.

\begin{table}[t]
    \centering
    \caption{Dataset to train and test our diarization models.}
    \subfloat[][Simulated datasets]{%
    \label{tbl:simulated_dataset}%
    \scalebox{\tablescale}{%
    \begin{tabular}{@{}llcccc@{}}
        \toprule
        Dataset & &\#Spk & \#Mixtures & $\beta$ & Overlap ratio (\%)\\\midrule
        \textbf{Train}
        &Sim1spk& 1 & 100,000&2&0.0\\
        &Sim2spk& 2 & 100,000&2&34.1\\
        &Sim3spk& 3 & 100,000&5&34.2\\
        &Sim4spk& 4 & 100,000&9&31.5\\\midrule
        \textbf{Test}&Sim1spk& 1 & 500 & 2&0.0\\
        &Sim2spk& 2 & 500&2&34.4\\
        &Sim3spk& 3 & 500&5& 34.7\\
        &Sim4spk& 4 & 500&9&32.0\\
        &Sim5spk& 5 & 500&13&30.7\\
        &Sim6spk& 6 & 500&17&29.9\\
        \bottomrule
    \end{tabular}%
    }
    }\\
    \vspace{5pt}
    \subfloat[][Real datasets]{%
    \label{tbl:real_dataset}%
    \scalebox{\tablescale}{%
    \begin{tabular}{@{}llcccc@{}}
        \toprule
        Dataset && Split&\#Spk & \#Mixtures & Overlap ratio (\%)\\\midrule
        \textbf{Adaptation}&CALLHOME \cite{callhome}&Part 1&2--7&249&17.0\\
        &DIHARD II \cite{ryant2019second}&dev&1--10&192&9.8\\
        &DIHARD III \cite{ryant2021third}&dev&1--10&254&10.7\\\midrule
        \textbf{Test}&CALLHOME \cite{callhome} &Part 2&2--6 & 250 & 16.7\\
        &DIHARD II \cite{ryant2019second}&eval &1--9 &194&8.9\\
        &DIHARD III \cite{ryant2021third}&eval&1--9&259&9.2\\
        \bottomrule
    \end{tabular}%
    }
    }
\end{table}

\subsection{Results}
\subsubsection{Simulated data}\label{sec:result_simu}
To show that the proposed method can deal with an unseen number of speakers, we first evaluated our model on the simulated datasets.
The results are shown in \autoref{tbl:result_simu}.
As an x-vector clustering baseline, we used the Kaldi recipe \footnote{\url{https://github.com/kaldi-asr/kaldi/tree/master/egs/callhome_diarization/v2}}, which resulted in poor performance (first row).
EEND-EDA performed well in matched conditions but DERs were rapidly degraded in mismatched conditions (second row).
We also show the results when a maximum of four attractors were used \ie, fifth and later attractors were ignored (third row).
DERs on the four-, five-, and six-speaker mixtures were improved by limiting the number of attractors (third row).
These results indicate that the fifth and subsequent attractors are of no use even if EDA estimates that the number of speakers is larger than four.

The proposed method trained using $\mathcal{L}_\text{local}$ in \autoref{eq:local_loss} performed better in mismatched conditions (fourth row), and the combined use of $\mathcal{L}_\text{global}$ and $\mathcal{L}_\text{local}$ in \autoref{eq:both_loss} further improved the DERs in both matched and mismatched conditions (fifth row).
However, local-attractor-based inference did not perform well in matched conditions, especially when the number of speakers was small (\eg, one or two).
This is because a small error in the estimated number of speakers (\eg, $\pm1$) can significantly degrade the DER.
By using global- and local-attractor-based estimation depending on the number of estimated speakers, the proposed method performed well in both matched and mismatched conditions (sixth row).
Increasing the number of Transformer encoders further improved the DERs in matched conditions, but the DERs in mismatched conditions were slightly degraded (seventh row).
This may be because the Transformers were overtrained to distinguish the seen number of speakers.
For comparison, we also show the results of the model that is trained on mixtures of at most five speakers in the last row, which is drawn from \cite{horiguchi2021encoderdecoder}.
It performed well on Sim5spk because it used five-speaker mixtures for training, but the DER sharply fell when six-speaker mixtures were input.
The proposed method achieved a comparative performance on Sim5spk even though it did not see five-speaker mixtures during training, and it also outperformed EEND-EDA on Sim6spk.

The confusion matrices for speaker counting are shown in \autoref{tbl:speaker_counting}.
We can clearly see that the proposed method could estimate the number of speakers in mismatched conditions with higher accuracies than the conventional EEND-EDA.
Indeed, EEND-EDA sometimes estimated the number of speakers as more than four, but considering the results in \autoref{tbl:result_simu}, the fifth and sixth attractors did not represent the fifth and sixth speakers.

\begin{table}[t]
    \centering
    \caption{DERs (\%) on the simulated datasets. Note that five- and six-speaker mixtures were not included in the training set. The best scores are \textbf{bolded} and the second best are \underline{underlined}. \#Blocks: the number of Transformer encoder blocks. Switch: use of global- and local-attractor-based inference depending on the estimated number of speakers.}
    \setlength{\tabcolsep}{3pt}
    \label{tbl:result_simu}
    \begin{threeparttable}
    \resizebox{\linewidth}{!}{%
    \begin{tabular}{@{}lccccccc@{}}
        \toprule
        &&\multicolumn{6}{c}{\#Speakers}\\\cmidrule(l){3-8}
        &&\multicolumn{4}{c}{matched}&\multicolumn{2}{c}{mismatched}\\\cmidrule(lr){3-6}\cmidrule(l){7-8}
        &\#Blocks&1&2&3&4&5&6\\\midrule
        X-vector clustering&N/A&37.42&7.74&11.46&22.45&31.00&38.62\\
        EEND-EDA \cite{horiguchi2020endtoend,horiguchi2021encoderdecoder} &4&\underline{0.15}&\textbf{3.19}&\underline{6.60}&9.26&23.11&34.97\\
        EEND-EDA \cite{horiguchi2020endtoend,horiguchi2021encoderdecoder} \tnote{\dag} &4&\underline{0.15}&\textbf{3.19}&\underline{6.60}&\underline{8.68}&22.43&33.28\\
        Proposed ($\mathcal{L}_\text{local}$)&4& 8.85&12.71&10.31&11.14&14.11&19.36\\
        Proposed ($\mathcal{L}_\text{local}+\mathcal{L}_\text{global}$)&4&2.84&10.21&7.54&9.08&\textbf{12.40}&\underline{18.03}\\
        Proposed ($\mathcal{L}_\text{local}+\mathcal{L}_\text{global}$, switch)&4&0.25&\underline{3.53}&6.79&8.98&\underline{12.44}&\textbf{17.98}\\
        Proposed ($\mathcal{L}_\text{local}+\mathcal{L}_\text{global}$, switch)&6&\textbf{0.09}&3.54&\textbf{5.74}&\textbf{6.79}&12.51&20.42\\\midrule
        EEND-EDA \cite{horiguchi2020endtoend,horiguchi2021encoderdecoder} \tnote{\ddag}  &4& 0.36&3.65&7.70&9.97&11.95&22.59\\
        \bottomrule
    \end{tabular}%
    }
    \begin{tablenotes}
        \footnotesize
        \item[\dag] At most four attractors were used.
    	\item[\ddag] Trained on Sim\{1,2,3,4,5\}spk. At most five attractors were used.
    \end{tablenotes}
    \end{threeparttable}
\end{table}
\begin{table}
    \caption{Confusion matrices for speaker counting.}
    \label{tbl:speaker_counting}
    \setlength{\tabcolsep}{3pt}
    \subfloat[][EEND-EDA \cite{horiguchi2020endtoend,horiguchi2021encoderdecoder}]{%
        \resizebox{0.48\linewidth}{!}{%
        \begin{tabular}{@{}cc|cccccc@{}}
            \toprule
            &&\multicolumn{6}{c}{Ref. \#Speakers}\\
            &&1&2&3&4&5&6\\\midrule
            \multirow{7}{*}{\rotatebox{90}{Pred. \#Speakers}}&1 & \textbf{500}&0&0&0&0&0\\
            &2 & 0&\textbf{482}&0&0&0&0\\
            &3 & 0&17&\textbf{435}&5&1&0\\
            &4 & 0&1&65&\textbf{447}&224&139\\
            &5 & 0&0&0&48&\textbf{268}&337\\
            &6 & 0&0&0&0&7&\textbf{24}\\
            &7+ & 0&0&0&0&0&0\\
            \bottomrule
        \end{tabular}%
    }%
    }
    \hfill
    \subfloat[][Proposed ($\mathcal{L}_\text{local}+\mathcal{L}_\text{global}$, switch, $\text{\#Blocks}=4$)]{%
        \resizebox{0.48\linewidth}{!}{%
        \begin{tabular}{@{}cc|cccccc@{}}
            \toprule
            &&\multicolumn{6}{c}{Ref. \#Speakers}\\
            &&1&2&3&4&5&6\\\midrule
            \multirow{7}{*}{\rotatebox{90}{Pred. \#Speakers}}&1 & \textbf{498}&0&0&0&0&0\\
            &2 & 2&\textbf{474}&0&0&0&0\\
            &3 & 0&25&\textbf{451}&17&2&1\\
            &4 & 0&1&33&\textbf{412}&78&30\\
            &5 & 0&0&10&62&\textbf{361}&183\\
            &6 & 0&0&6&7&47&\textbf{229}\\
            &7+ & 0&0&0&2&12&57\\
            \bottomrule
        \end{tabular}%
    }%
    }
\end{table}

\subsubsection{CALLHOME}
\begin{table}[t]
    \centering
    \caption{DERs (\%) on CALLHOME dataset. The switching strategy was used for the proposed method.}
    \setlength{\tabcolsep}{3pt}
    \label{tbl:results_callhome}
    \begin{threeparttable}
    \resizebox{\linewidth}{!}{%
    \begin{tabular}{@{}lccccccc@{}}
        \toprule
        &&\multicolumn{5}{c}{\#Speakers}\\\cmidrule(l){3-7}
        Method&\#Blocks&2&3&4&5&6&All\\\midrule
        VBx \cite{landini2022bayesian} \tnote{\dag}&N/A& 9.44&13.89&16.05&\textbf{13.87}&24.73&13.28\\
        SC-EEND \cite{fujita2020neural}&4&9.57&14.00&21.14&31.07&37.06&15.75\\
        EEND-EDA \cite{horiguchi2020endtoend,horiguchi2021encoderdecoder}&4&7.83&12.29&17.59&27.66&37.17&13.65\\
        EEND-vector clust. \cite{kinoshita2021advances} &6&7.94&11.93&16.38&\underline{21.21}&\underline{23.10}&12.49\\
        Proposed ($\mathcal{L}_\text{local}+\mathcal{L}_\text{global}$, switch)&4&\textbf{6.94}&\textbf{11.42}&\underline{14.49}&29.76&24.09&\underline{11.92}\\
        Proposed ($\mathcal{L}_\text{local}+\mathcal{L}_\text{global}$, switch)&6&\underline{7.11}&\underline{11.88}&\textbf{14.37}&25.95&\textbf{21.95}&\textbf{11.84}\\\bottomrule
    \end{tabular}%
    }
    \begin{tablenotes}
        \footnotesize
        \item[\dag] Oracle speech segments were used.
    \end{tablenotes}
    \end{threeparttable}
\end{table}

We also evaluated our method on the CALLHOME dataset.
As comparison methods, the state-of-the-art x-vector-based method (VBx) \cite{landini2022bayesian} and several EEND-based methods that can deal with a flexible number of speakers \cite{fujita2020neural,horiguchi2020endtoend,horiguchi2021encoderdecoder,kinoshita2021advances} were adopted.
Note that VBx used the oracle speech segments while EEND-based methods estimated speech activities from the input audio.

\autoref{tbl:results_callhome} shows the number-of-speakers-wise DERs.
Our method achieved \SI{11.92}{\percent} and \SI{11.84}{\percent} DERs by using four- and six-stacked Transformer encoders as a backbone, respectively, which were better than the conventional methods.
Compared with EEND-EDA, the proposed method improved the DERs, especially when the number of speakers was large.

\subsubsection{DIHARD II \& III}
\begin{table}[t]
    \centering
    \caption{DERs / JERs (\%) on DIHARD II dataset.}
    \setlength{\tabcolsep}{3pt}
    \label{tbl:results_dihard2}
    \resizebox{\linewidth}{!}{%
    \begin{tabular}{@{}lcccc@{}}
        \toprule
        &&\multicolumn{2}{c}{\#Speakers}\\\cmidrule(l){3-4}
        Method&\#Blocks&$\leq4$&$\geq5$&All\\\midrule
        VBx + overlap handling \cite{landini2020but} & N/A&\textbf{21.34} / 43.00&\textbf{39.85} / \textbf{57.40}&\textbf{27.11} / \textbf{49.07}\\
        EEND-EDA \cite{horiguchi2020endtoend,horiguchi2021encoderdecoder} &4&22.09 / 40.70&47.66 / 71.49&30.07 / 53.69\\
        Proposed ($\mathcal{L}_\text{local}+\mathcal{L}_\text{global}$, switch)&4&22.24 / \underline{40.47}&44.92 / 69.34&29.31 / 52.64\\
        Proposed ($\mathcal{L}_\text{local}+\mathcal{L}_\text{global}$, switch)&6&\underline{21.40} / \textbf{37.87}&\underline{43.62} / \underline{68.09}&\underline{28.33} / \underline{50.62}\\\bottomrule
    \end{tabular}%
    }
    \vspace{10pt}
    \caption{DERs / JERs (\%) on DIHARD III dataset.}
    \setlength{\tabcolsep}{3pt}
    \label{tbl:results_dihard3}
    \resizebox{\linewidth}{!}{%
    \begin{tabular}{@{}lcccc@{}}
        \toprule
        &&\multicolumn{2}{c}{\#Speakers}\\\cmidrule(l){3-4}
        Method&\#Blocks&$\leq4$&$\geq5$&All\\\midrule
        VBx + overlap handling \cite{horiguchi2021hitachi} & N/A&16.38 / 29.43&\textbf{42.51} / \textbf{53.47} & 21.47 / \textbf{37.83}\\
        EEND-EDA \cite{horiguchi2020endtoend,horiguchi2021encoderdecoder} &4&15.55 / 27.23&48.30 / 71.76 & 21.94 / 42.79\\
        Proposed ($\mathcal{L}_\text{local}+\mathcal{L}_\text{global}$, switch)&4&\underline{14.39} / \underline{25.85} & 44.32 / \underline{68.06} & 20.23 / 40.60\\
        Proposed ($\mathcal{L}_\text{local}+\mathcal{L}_\text{global}$, switch)&6&\textbf{13.64} / \textbf{24.60} & \underline{43.67} / 69.33 & \textbf{19.49} / \underline{40.23}\\\bottomrule
    \end{tabular}%

    }
\end{table}

Finally, we evaluated our method on the DIHARD II and III datasets, which contain mixtures of at most nine speakers.
We used VBx-based systems by BUT \cite{landini2020but} and the Hitachi-JHU team \cite{horiguchi2021hitachi} as clustering-based baselines.
Because they are challenge submissions, each of them is carefully tuned to concatenate multiple modules including speech activity detection, speech dereverberation, x-vector extraction, probabilistic linear discriminant analysis scoring, and overlap detection and assignment.
EEND-EDA was also used as a comparison method.

Tables \ref{tbl:results_dihard2} and \ref{tbl:results_dihard3} show DERs and JERs on the DIHARD II and III datasets, respectively.
The proposed method improved DERs especially when the number of speakers was larger than four compared to EEND-EDA, and achieved \SI{28.33}{\percent} DER on the DIHARD II dataset and \SI{19.49}{\percent} DER on the DIHARD III dataset with six-stacked Transformer encoders.
These DERs are close to those of carefully tuned challenge submissions, and in particular even better on the DIHARD III dataset.

\section{Conclusion}
In this paper, we proposed a neural diarization method based on global and local attractors.
An input sequence of acoustic features is first converted into a sequence of frame-wise embeddings using stacked Transformer encoders and then divided into short subsequences.
We calculate the diarization results based on speaker-wise local attractors for each subsequence, followed by unsupervised clustering based on the local attractors to find the optimal correspondence between subsequences.
The experimental results demonstrated the effectiveness of our method, especially when the number of speakers is large.
Future work will include an online extension of the proposed method using online clustering methods.

\bibliographystyle{IEEEbib}
\bibliography{mybib}
\end{document}